\documentclass[journal]{IEEEtran}

\ifCLASSINFOpdf
\else
   \usepackage[dvips]{graphicx}
\fi
\usepackage{url}

\usepackage{graphicx}
\usepackage{multirow}
\usepackage{enumerate}
\usepackage{subfig}
\usepackage{amsthm,amsmath,amssymb}
\usepackage{mathrsfs}
\usepackage{dutchcal}
\usepackage{cite}
\usepackage{lipsum,amsmath}
\usepackage{stfloats}
\usepackage{diagbox}
\usepackage{balance}
\usepackage{color}
\usepackage{bm}
\usepackage{amsmath}
\begin{document}

\title{ Graph Fourier Transform based Audio Zero-watermarking }

\author{Longting Xu, \IEEEmembership{Member, IEEE}, Daiyu Huang, Syed Faham Ali Zaidi, Abdul Rauf, and \\Rohan Kumar Das, \IEEEmembership{Senior Member, IEEE}
\thanks{This work is supported by the National Natural Science Foundation of China (No.62071242, No.62001100), the Fundamental Research Funds for the Central Universities (No.2232019D3-52), and Shanghai Sailing Program (No.19YF1402000).}
\thanks{L. Xu, D. Huang, S. Zaidi and A. Rauf are with the College of Information Science and Technology, Donghua University, Shanghai 200000, China. (e-mail: xlt@dhu.edu.cn).}
\thanks{R. K. Das, is with Fortemedia Singapore, Singapore 138589. (e-mail: ecerohan@gmail.com).}
}

\markboth{Journal of \LaTeX\ Class Files, Vol. 14, No. 8, August 2015}
{Shell \MakeLowercase{\textit{et al.}}: Bare Demo of IEEEtran.cls for IEEE Journals}
\maketitle

\begin{abstract}

The frequent exchange of multimedia information in the present era projects an increasing demand for copyright protection. In this work, we propose a novel audio zero-watermarking technology based on graph Fourier transform for enhancing the robustness with respect to copyright protection. In this approach, the combined shift operator is used to construct the graph signal, upon which the graph Fourier analysis is performed. The selected maximum absolute graph Fourier coefficients representing the characteristics of the audio segment are then encoded into a feature binary sequence using K-means algorithm. Finally, the resultant feature binary sequence is XOR-ed with the watermark binary sequence to realize the embedding of the zero-watermarking. The experimental studies show that the proposed approach performs more effectively in resisting common or synchronization attacks than the existing state-of-the-art methods.

\end{abstract}


\IEEEpeerreviewmaketitle

\section{Introduction}
\IEEEPARstart{T}{he} rapid development of the Internet and multimedia technology has made the exchange of multimedia information reach an unprecedented depth. However, at the same time piracy and tampering have become rampant for it \cite{Non-Linear,blind}. Digital watermarking technology emerged as one of its solutions and has received widespread attention in the recent years \cite{HE,PN}. While various multimedia data (including audio, image, and video) require copyright protection, this paper focuses on protection of audio data.


Audio zero-watermarking technology is a promising technology for audio copyright protection due to its excellent imperceptibility, while there is potential to improve its robustness \cite{8577943}. The zero-watermarking technology does not modify the data of the host audio signal, but constructs watermark information based on its content characteristics \cite{V-map}. Based on different techniques for exploring stable content characteristics, audio zero-watermarks can be broadly grouped into two categories: time domain vector mapping and transform domain approach. In general, the singular value decomposition (SVD) is used to seek stable characteristics based on the time domain. The largest singular value with stability is adopted to represent audio characteristics and the repetition of audio content makes the SVD based zero-watermark method to effectively resist serious synchronization attacks \cite{CONTENT, beats}.

Recently, researchers have been also studied watermarking technology based on the transform domain, including discrete cosine transform (DCT) \cite{PN,HE,DCT--HT,imperceptible,MP3}, discrete wavelet transform (DWT) \cite{DWT,DWT-dsss}, Fourier transform (FT) \cite{dft,stft} and linear prediction cepstrum coefficients (LPCC) \cite{lpcc}. Specifically, DCT domain-based technologies \cite{PN,HE,DCT--HT,imperceptible,MP3} apply the DCT to the host audio signal to obtain a set of audio segments. However, these technological innovations lie in the selection of audio segment characteristics. The DWT domain based technologies \cite{DWT,DWT-dsss} are similar to the DCT approach, which perform DWT on the host signal to select the stable characteristics of the audio segment. Furthermore, the phase information based on FT is also commonly used to characterize the characteristics of audio segments \cite{dft,stft}.

The technologies based on amplitude information (such as DCT, DWT, or a combination of multiple transform domains \cite{dhar2017blind,DWT-DCT,lei2016audio}) to characterize audio segment characteristics can effectively resist common attacks (such as noise, filtering and re-sampling attacks, etc.), whereas they generally do not have the ability to resist harsh synchronization attacks (such as time scale modify (TSM), cropping attacks). On the contrary, the technologies based on phase information of FT can resist synchronization attacks better than the technologies based on amplitude information. However, they are not very effective to resist common attacks.

To further enhance the robustness against various nature of attacks, a novel zero-watermarking technology for audio based on the graph Fourier transform (GFT) \cite{gft} is proposed in this work. The emerging graph signal processing (GSP) \cite{limit,graph} technology has been used in speech processing to express the structural relationship of speech sample data points. The graph topology constructed by the potential relationship between data points can determine the graph Fourier basis \cite{gss}, and GFT can further analyze the characteristics of the graph signal in the graph frequency domain \cite{gft}. Progress has been made in using GFT to watermark unstructured data, such as point clouds \cite{3D,GS,2020A} and graphic data \cite{sensor}. Here, we adopt GFT to transform the speech signal from the graph domain to the graph frequency domain for stabilizing characteristics of audio segments. We then encode all the selected graph Fourier coefficients that can characterize the audio segments to achieve zero-watermark embeddings. We evaluate our proposed method against various methods to show effectiveness against both common and synchronization attacks.


\section{Graph Spectrum of Audio Signal}
\label{sec:GFT}
In this section, we describe the process to map the audio signal in the time domain to the graph domain and then the details to convert the graph signal to the graph frequency domain for further analyzing its characteristics.

\subsection{Graph domain mapping of audio signal}

\subsubsection{Basic concept of graph signal}
In order to map the audio signal $\boldsymbol{x}$ in the time domain to the graph audio signal $\boldsymbol{y}$ in the graph domain, it is necessary to exploit the graph in the GSP. The graph is composed  of vertices, edges connecting the vertices, and edge weights. Mathematically, the graph can be expressed as, $\boldsymbol{\mathcal{G}}=(\boldsymbol{\mathcal{V}},\boldsymbol{\mathcal{E}},\boldsymbol{\mathcal{W}})$, where $\boldsymbol{\mathcal{V}}$ represents the vertex set, $\boldsymbol{\mathcal{E}}$ represents the edge set, and $\boldsymbol{\mathcal{W}}$ represents the weight set \cite{graph}.

In order to utilize GSP for processing audio signals, the time domain signal $\boldsymbol{x}$ is first divided into frames, and then each frame is mapped. Assuming that $\boldsymbol{x}$ is divided into $M$ frames with $N$ sampling points, one of the frames can be expressed as $\boldsymbol{{x_m}} = {[{x_m}_1,{x_m}_2,...,{x_m}_N]^T}$ and $m=1,2,...,M$. Given a graph, $\boldsymbol{y_m}$ can be expressed as a graph signal, which is defined as a mapping as follows.
\begin{equation}
\mathcal{S}:\boldsymbol{{x_m}} \to \boldsymbol{{y_m}}{\rm{ }},
\label{eq:map}
\end{equation}
where, $\boldsymbol{{y_m}}={[{y_m}_1,{y_m}_2,...,{y_m}_N]^T} $ indexed by $\boldsymbol{\mathcal{G}}=(\boldsymbol{\mathcal{V}},\boldsymbol{\mathcal{E}},\boldsymbol{\mathcal{W}})$ is a one-to-one mapping value of $\boldsymbol{x_m}$. Each element of $\boldsymbol{y_m}$ represents the intensity at a vertex in the corresponding graph and each vertex corresponds to a sampling point in the time domain. The corresponding graph $\boldsymbol{\mathcal{G}}$ describes the relationship among the vertices and can be written in detail as given in Equation (\ref{eq:graph}).
\begin{equation}
\begin{array}{l}
\boldsymbol{\mathcal{V}} = {[{\mathcal{v}_1},{\mathcal{v}_2},...,{\mathcal{v}_N}]^T},\\
\boldsymbol{\mathcal{E}} = {\left\{ {{\mathcal{e}_{ij}} \in \{ 0,1\} } \right\}_{i = 1,2,...,N,}}_{j = 1,2,...,N} \in {\mathbb{R}^{N \times N}},\\
\boldsymbol{\mathcal{W}} = {\left\{ {{\mathcal{w}_{ij}}} \right\}_{(i,j) \in \mathcal{E}}} \in {\mathbb{R}^{N \times N}}.
\end{array}
\label{eq:graph}
\end{equation}
Here, ${\mathcal{e}_{ij}} = 0$ indicates that there is no edge connection between vertex $\mathcal{v}_i$ and $\mathcal{v}_j$, otherwise ${\mathcal{e}_{ij}} = 1$. The ${\mathcal{w}_{ij}}$ represents the weight of the edge between $ \mathcal{v}_i$ and $\mathcal{v}_j$. The general weight matrix can be represented by the graph Laplacian matrix $\boldsymbol{\mathcal{L}}$ or the graph adjacency matrix $\boldsymbol{\mathcal{A}}$ \cite{L2016,A2018}. Among them, $\boldsymbol{\mathcal{L}}$ is only applicable to undirected graphs, while the $\boldsymbol{\mathcal{A}}$ is not \cite{limit}. Considering speech signal is a time series with obvious temporal relevance, directional weights can exactly represent the relationship between speech time sampling points. Therefore, this work adopts $\boldsymbol{\mathcal{A}}$ as $\boldsymbol{\mathcal{W}}$ and the value of $\boldsymbol{\mathcal{A}}$'s elements are 0 or 1 to achieve the purpose of focusing only on whether there is a connection between the vertices.

\subsubsection{Construction of graph audio signal}
In this work, the combined graph $k$-shift operator $\boldsymbol{{\Gamma _k}}$ is used to construct $\boldsymbol{\mathcal{A}}$ to obtain the graph speech signal. According to the above analysis $\boldsymbol{\mathcal{A}}$, $\boldsymbol{\mathcal{A}}$ is equivalent to $\boldsymbol{\mathcal{W}}$ and $\boldsymbol{\mathcal{E}}$ as a binary matrix, the graph can be redefined as $\boldsymbol{{\mathcal{G}_{{\Gamma _k}}}} = (\boldsymbol{\mathcal{V}},\boldsymbol{{\Gamma _k}},\boldsymbol{{\Gamma _k}})$. $\boldsymbol{{\Gamma _k}}$ is defined as
\begin{equation}
\boldsymbol{{\Gamma _k}} = \sum\limits_{t = 0}^{k - 1} {\boldsymbol{{{\gamma _t}}},k = 1,2,...,}
\label{eq:combine}
\end{equation}
where $\boldsymbol{{\gamma _t}} \in {\mathbb{R}^{N \times N}}(t = 0,1,...,)$ is a binary matrix and which denotes a $t$-shift operator. The element ${\gamma _{ij}}$ of $\boldsymbol{{\Gamma _k}}$ satisfies the condition
\begin{equation}
{\gamma _{ij}} = \left\{ {\begin{array}{l}
	{1,if(j - i)\bmod N = 0,...,k-1}\\
	{0,else}
	\end{array}} \right..
\label{eq:one}
\end{equation}

Obviously, when $k=1$, $\boldsymbol{{\Gamma _1}} = \boldsymbol{{\gamma _0}}$ is a unit matrix which implies that the signal was not shifted. The graph signal $\boldsymbol{{y_o}}$ obtained after implementing $\boldsymbol{\Gamma _k}$ on the time domain signal $\boldsymbol{{y_i}}$ can be expressed as
$\boldsymbol{{y_o}} = \boldsymbol{{\Gamma _k}} \cdot \boldsymbol{{y_i}}$.

\subsection{Spectrum of audio signal in graph frequency domain}

With the aid of the adjacency matrix $\boldsymbol{\mathcal{A}}$, the graph domain signal can be converted to the graph frequency domain. The specific method performs SVD on $\boldsymbol{\mathcal{A}}$ to obtain the singular value decomposition of $\boldsymbol{\mathcal{A}}$. We have $\boldsymbol{\mathcal{A}} = \boldsymbol{\mathcal{Q}}\boldsymbol{\Sigma} {\boldsymbol{\mathcal{Q}}^{ - 1}}$, where $\boldsymbol{\mathcal{Q}} = [\boldsymbol{{\varepsilon _1}},\boldsymbol{{\varepsilon _2}},...,\boldsymbol{{\varepsilon _N}}] \in {\mathbb{R}^{N \times N}} $ formed by $N$ eigenvectors of $\boldsymbol{\mathcal{A}}$ and $\boldsymbol{\Sigma } = [\boldsymbol{{\zeta _1}},\boldsymbol{{\zeta _2}},...,\boldsymbol{{\zeta _N}}] \in {\mathbb{R}^{N \times N}}$ with $N$ eigenvectors as
the main diagonal of $\boldsymbol{\mathcal{A}}$. The column $\boldsymbol{\varepsilon _t}$ of $\boldsymbol{\mathcal{Q}}$ represents the spectral components at the corresponding graph frequency $\boldsymbol{\zeta _t}$.
	
Since $\boldsymbol{\mathcal{A}}$ here is a row trapezoidal matrix with full row rank, this will result in $N$ linearly independent eigenvectors. Correspondingly, $\boldsymbol{\mathcal{Q}}$ is invertible, the graph Fourier matrix $\boldsymbol{\mathcal{F}}$ can be defined as follows \cite{gft}.
\begin{equation}
\begin{aligned}
\boldsymbol{\mathcal{F}} &= {\boldsymbol{\mathcal{Q}}^{ - 1}}  = {[\boldsymbol{{\varepsilon _1}},\boldsymbol{{\varepsilon _2}},...,\boldsymbol{{\varepsilon _N}}]^{ - 1}}\\
&= [{\boldsymbol{\mathcal{f}}_1},{\boldsymbol{\mathcal{f}}_2},...,{\boldsymbol{\mathcal{f}}_N}].
\end{aligned}
\label{eq:F}
\end{equation}

The graph spectrum $\boldsymbol{\tilde y}$ obtained after performing the GFT on the graph signal $\boldsymbol{y}$ can be expressed as
 \begin{equation}
\begin{aligned}
\boldsymbol{\tilde y} &= \boldsymbol{\mathcal{F}} \cdot \boldsymbol{y} = {\boldsymbol{\mathcal{Q}}^{ - 1}} \cdot \boldsymbol{y}\\
&= {[{\boldsymbol{\mathcal{f}}_1}\boldsymbol{y},{\boldsymbol{\mathcal{f}}_2}\boldsymbol{y},...,{\boldsymbol{\mathcal{f}}_N}\boldsymbol{y}]^T}\\
&= {[{{\boldsymbol{\tilde y}}_{{\boldsymbol{\mathcal{f}}_1}}},{{\boldsymbol{\tilde y}}_{{\boldsymbol{\mathcal{f}}_2}}},...,{{\boldsymbol{\tilde y}}_{{\boldsymbol{\mathcal{f}}_N}}}]^T},
\end{aligned}
\label{eq:gft}
 \end{equation}
where ${\boldsymbol{\tilde y}}_{{\mathcal{f}_t}}$ represents the graph Fourier coefficients at the corresponding graph frequency $\boldsymbol{\zeta _t}$. In addition, combining the equations (\ref{eq:F}) and (\ref{eq:gft}) to note that GFT is essentially a simple matrix operation process, and it is a method without latency. 


\begin{figure}[t]
	\centering
	\includegraphics[width=0.5\textwidth]{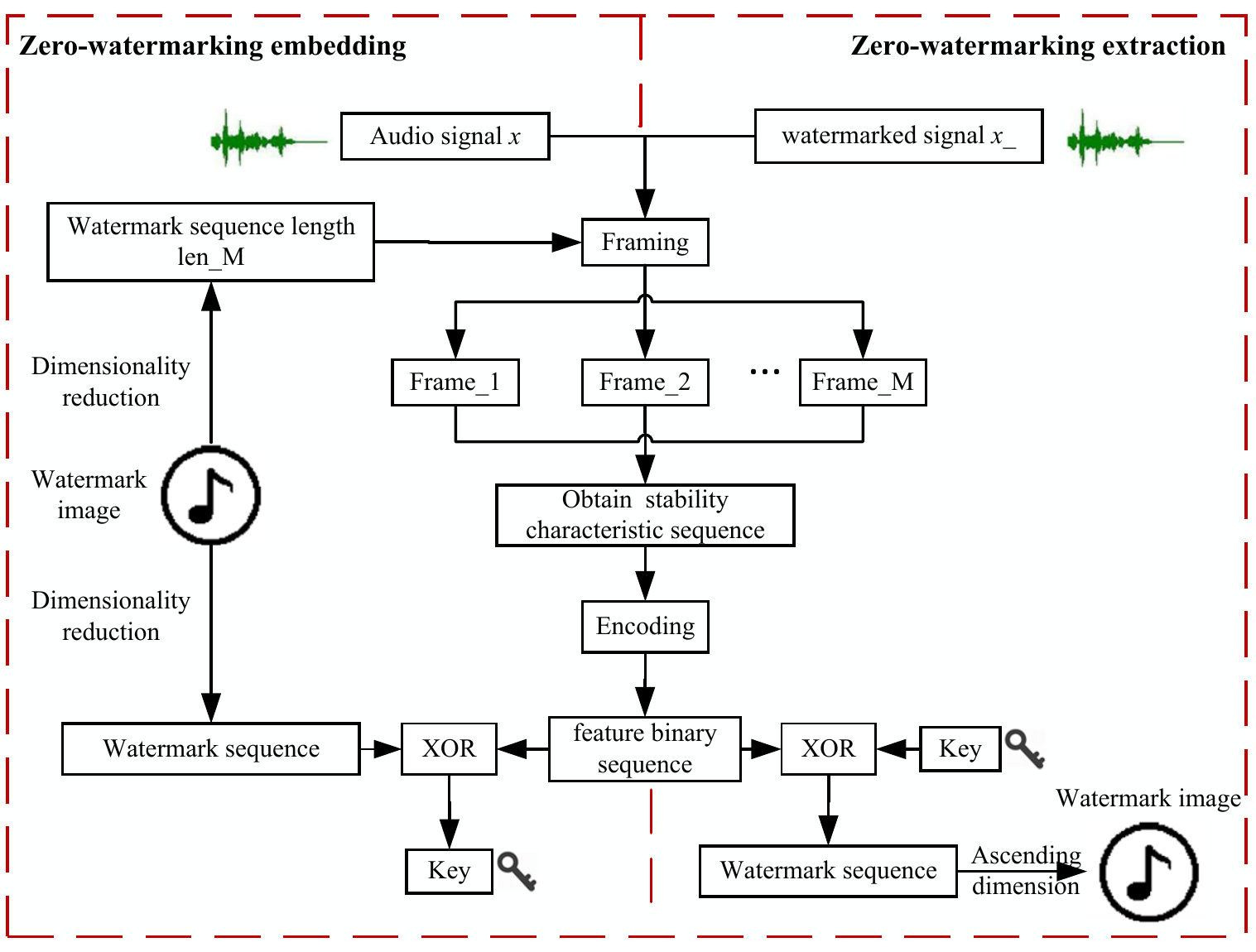}	
	\caption{Proposed zero-watermarking scheme: left: zero-watermark embedding; right: zero-watermark extraction.}
	\label{fig:em-ex}	
\end{figure}

\section{Proposed GFT based audio zero-watermarking}
\label{sec:scheme} 

The proposed framework performs zero-watermark processing in the GFT domain. Fig. \ref{fig:em-ex} shows the flow diagram of the zero-watermarking generation and extraction process. The zero-watermarking embedding and extraction have some common processes that include framing, constructing graph signal, GFT and encoding. We note that their XOR process is slightly different.

\subsection{The common processes}

%
%

\begin{table*}[b]
		\centering
		\caption{Robustness comparison of the proposed scheme and the baselines under common attacks, where $**/**$ indicates average metrics BER/NC and the bold mark indicates the best number across all the schemes under each attack.}
		\label{tab:attack_c}
		\resizebox{\textwidth}{!}{
			\begin{tabular}{c|cccccccc}
 \hline
  \multirow{2}{*}{\diagbox{Schemes}{Attacks}}& \multicolumn{8}{c}{Metrics (BER/NC)}\\
  \cline{2-9}
  & AWGN (10dB) & AWGN (20dB) & LPH & Re-sampling & Re-quantization& MP3& Amplitude (1.5)& Amplitude (2)\\
   \hline
  Our& \textbf{0.0308}/\textbf{0.9786} & \textbf{0.0083}/\textbf{0.9942} & 0.0130/0.9910& 0.0017/0.9988 & \textbf{0.0275}/\textbf{0.9810}& 0.0055/0.9961& 0.0003/0.9996& 0.0004/0.9995 \\	
  
  DWT \cite{DWT} & 0.0764/0.9467 & 0.0350/0.9758 & \textbf{0.0097}/\textbf{0.9932} & 0.0032/0.9977 & 0.0729/0.9492& 0.0164/0.9887& \textbf{0}/\textbf{1}& \textbf{0}/\textbf{1}\\
  STFT\cite{stft} & 0.2542/0.8141 & 0.2432/0.8227 & 0.2181/0.8419 & 0.1521/0.8916 & 0.2326/0.8307& 0.2091/0.8490& 0/0.9999& 0/1\\
  DWT-DCT \cite{DWT-DCT} & 0.1501/0.8934 &0.0680/0.9527 & 0.0376/0.9740 &0.0155/0.9893 & 0.0794/0.9445& 0.0240/0.9834& 0/1& 0/1\\
  DWT-DCT-SVD \cite{lei2016audio}& 0.0326/0.9774 & 0.0093/0.9935 & 0.0146/0.9899 & \textbf{0.0009}/\textbf{0.9993} & 0.1002/0.9295& \textbf{0.0036}/\textbf{0.9975}& 0.5333/0.5669& 0.5255/0.5776\\

   \hline		
			\end{tabular}}
	\end{table*}

\subsubsection{Framing}
According to the length $M$ of the watermark sequence obtained by watermark image dimensionality reduction, the audio signal $x$ is evenly divided into $M$ non-overlapping frames. The length of each frame is represented by $N$. Hence, we have $N = floor(x\_len/M)$
, where $x\_len$ denotes the length of the audio signal $x$.

\subsubsection{Constructing graph signal}

Once the time domain signal is framed, one of the frames can be expressed as $\boldsymbol{x}^{(m)} = {[{x_1}^{(m)},{x_2}^{(m)},...,{x_N}^{(m)}]^T}$ and $m=1,2,...,M$. By performing the combined graph $k$-shift operator ${\boldsymbol{\Gamma _k}}$ on $\boldsymbol{x}^{(m)}$, we can obtain the graph signal $\boldsymbol{y}^{(m)}$.
\begin{equation}
{\boldsymbol{y}^{(m)}} = {\boldsymbol{\Gamma _k}} \cdot {\boldsymbol{x}^{(m)}}.
\label{graph}
\end{equation}

\subsubsection{GFT}
Based on applying SVD on ${\boldsymbol{\Gamma _k}}$, the GFT base ${\mathcal{F}_{{\boldsymbol{\Gamma _k}}}}$ can be obtained. The graph spectrum coefficients ${{\boldsymbol{\tilde y}}_{{\Gamma _k}}}$ of the graph signal $\boldsymbol{y}^{(m)}$ can be then obtained by the formula (\ref{eq:gftt}).
\begin{equation}
{{\boldsymbol{\tilde y}}_{{\boldsymbol{\Gamma _k}}}} = {\boldsymbol{\mathcal{F}}_{{\boldsymbol{\Gamma _k}}}}{\boldsymbol{y}^{(m)}}.
\label{eq:gftt}
\end{equation}

Considering graph spectrum is mainly concentrated at lower frequencies, and when $k$ is small, the spectrum is relatively stable \cite{gss}. In addition, a larger $k$ would cause a higher amount of calculation. We mainly discuss the case when $k=3$ in this work.

\begin{figure}[t]
	\centering
	\includegraphics[width=0.45\textwidth]{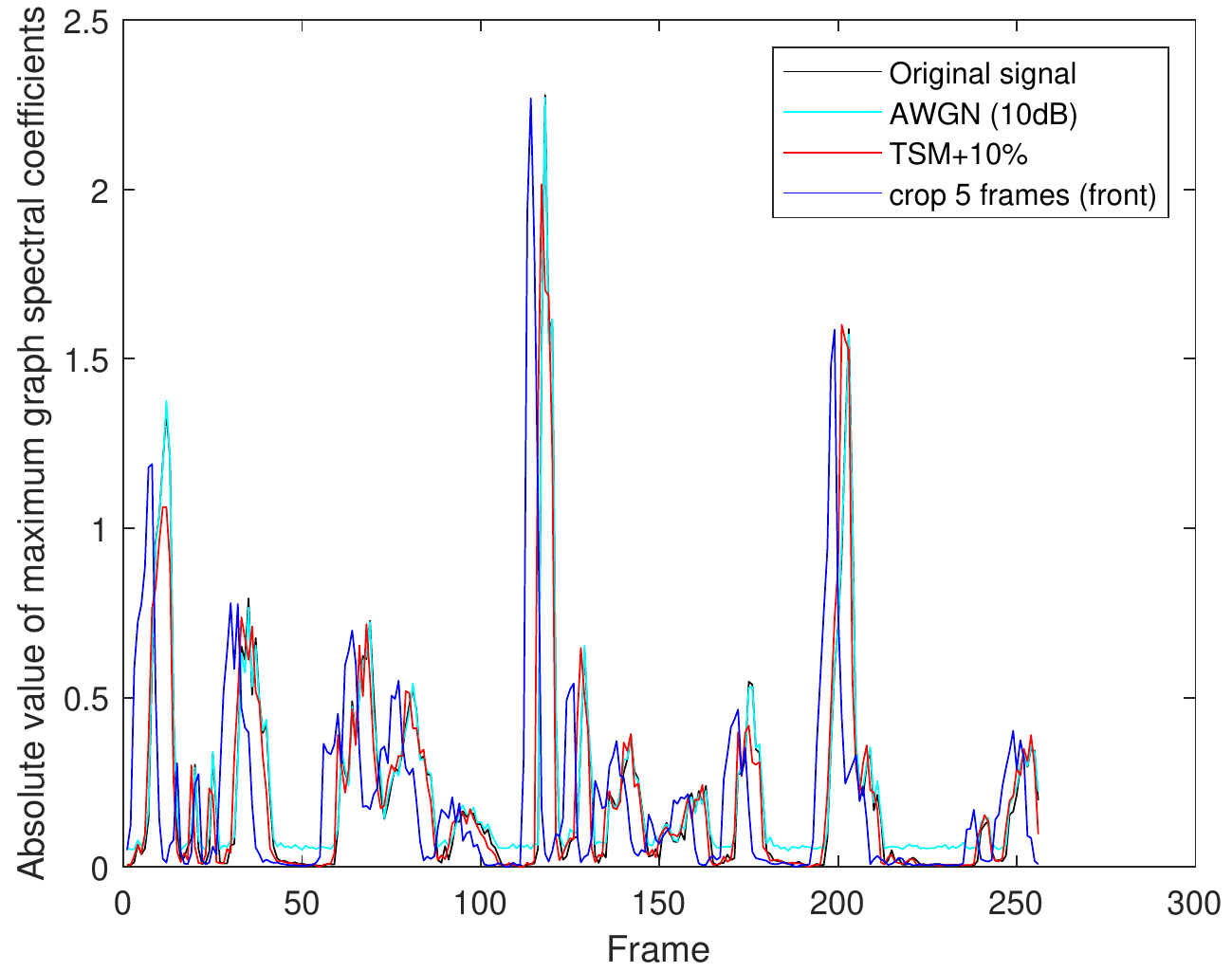} 	
	\caption{the maximum absolute graph spectral coefficient values under different attacks.}
	\label{fig:max}
\end{figure}

\subsubsection{Encoding}
For obtaining the stability characteristic sequence of the audio segment, we analyzed absolute value of the maximum spectral coefficient of the segments in audio with a duration of 22 seconds. Fig. \ref{fig:max} shows the maximum absolute graph spectral coefficient values under different attacks of the first 256 frames in 1024 frames. From Fig. \ref{fig:max}, it can be clearly observed that these values have undergone some changes after these attacks, but the trends are relatively stable. Therefore, these values can be used to represent the feature sequence $\boldsymbol{F}$ of each frame to resist attacks.

In order to obtain the feature binary sequence $\boldsymbol{B}$ of each audio segment, the K-means clustering algorithm is used to divide the feature sequence $\boldsymbol{F}$ into two categories, which are coded as 0 and 1, respectively.

\subsection{The different processes}
\subsubsection{XOR of zero-watermarking embedding}
The obtained signal feature binary sequence is XOR-ed with the watermark sequence to obtain the watermark key $\boldsymbol{K}$ as follows.
\begin{equation}
\boldsymbol{K} = \left\{ {k\left( m \right) = \boldsymbol{B}(m) \oplus \boldsymbol{{W}}(m)\left| {1 \le m \le M} \right.} \right\},
\end{equation}
where $\boldsymbol{W}(m)$ is the pixel point value of the binary image.
\subsubsection{XOR of zero-watermarking extraction}
The watermarked signal feature binary sequence $\boldsymbol{B'}$ is XOR-ed with the watermark key $\boldsymbol{K}$ to obtain the watermark sequence $\boldsymbol{W'}$ as follows.
\begin{equation}
\boldsymbol{W'} = \left\{ {w\left( m \right) = \boldsymbol{B'}(m) \oplus \boldsymbol{{K}}(m)\left| {1 \le m \le M} \right.} \right\}.
\end{equation}

\begin{figure}[h]
		\centering
		\subfloat[]{	\includegraphics[width=1cm]{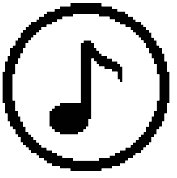}}
		\subfloat[]{	\includegraphics[width=1cm]{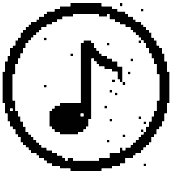}}
		\subfloat[]{	\includegraphics[width=1cm]{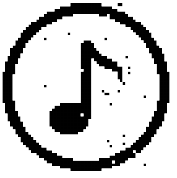}}
		\subfloat[]{	\includegraphics[width=1cm]{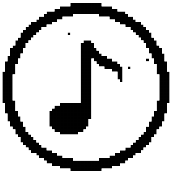}}
		\subfloat[]{	\includegraphics[width=1cm]{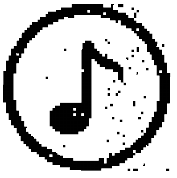}}
		\subfloat[]{	\includegraphics[width=1cm]{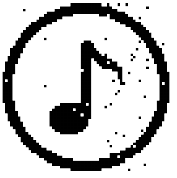}}
        \subfloat[]{	\includegraphics[width=1cm]{images/20.pdf}}	\\
        \subfloat[]{	\includegraphics[width=1cm]{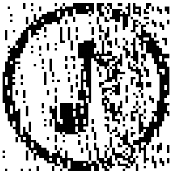}}
        \subfloat[]{	\includegraphics[width=1cm]{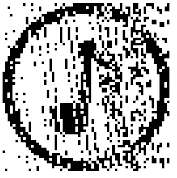}}
		\subfloat[]{	\includegraphics[width=1cm]{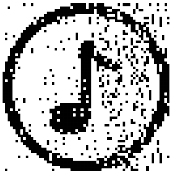}}
        \subfloat[]{	\includegraphics[width=1cm]{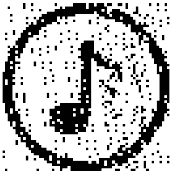}}
		\subfloat[]{	\includegraphics[width=1cm]{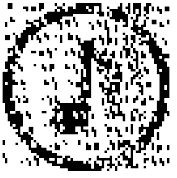}}
		\subfloat[]{	\includegraphics[width=1cm]{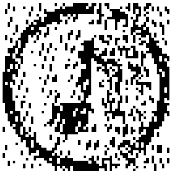}}
		\subfloat[]{	\includegraphics[width=1cm]{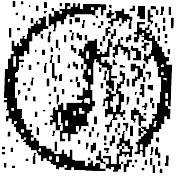}}
		\caption{Extracted watermark images of the proposed scheme under attacks: (a) Original image;
(b) AWGN (10dB); (c) LPH; (d) Re-sampling; (e) MP3; (f) Re-quantization;
(g) Amplitude (2 times); (h) TSM+1\%; (i) TSM+10\%; (j) TSM-1\%; (k) TSM-10\%;
(l) Cropping 5 frames (front); (m) Cropping 10 frames (front); (n) Cropping 20 frames (front)}
		\label{fig:water}	
	\end{figure}

Finally, the obtained watermark binary sequence can be restored to a watermark image by increasing the dimension, as shown in Fig. \ref{fig:water}.

\begin{table*}[t]
		\centering
		\caption{Robustness comparison of the proposed scheme and the baselines under synchronization attacks, where $**/**$ indicates average metrics BER/NC and the bold mark indicates the best number across all the schemes under each attack.}
		\label{tab:attack_cc}
		\resizebox{\textwidth}{!}{
			\begin{tabular}{c|cccccccccc}
 \hline
  \multirow{2}{*}{\diagbox{Schemes}{Attacks}}& \multicolumn{10}{c}{Metrics (BER/NC)}\\
  \cline{2-11}
  & TSM+1\% & TSM+10\% & TSM-1\% & TSM-10\% & Crop (5 front)& Crop (10 front)& Crop (20 front)& Crop (5 back)&Crop (10 back)&Crop (20 back)\\
   \hline
    Our& \textbf{0.1720}/\textbf{0.8760}&\textbf{0.1767}/\textbf{0.8725}&\textbf{0.1162}/\textbf{0.9175} & \textbf{0.1328}/\textbf{0.9052} & \textbf{0.1830}/\textbf{0.8675}& \textbf{0.2122}/\textbf{0.8452}&\textbf{0.2427}/\textbf{0.8211}& \textbf{0.1854}/\textbf{0.8648}&\textbf{0.2386}/\textbf{0.8243}&\textbf{0.2658}/\textbf{0.8028} \\	
	DWT \cite{DWT} & 0.5096/0.5966 & 0.4991/0.6058 & 0.4455/0.6536 & 0.4822/0.6211 & 0.4958/0.6093& 0.4997/0.6054& 0.4982/0.6071& 0.1975/0.8443&0.4929/0.6119&0.4967/0.6084\\	
   STFT\cite{stft} & 0.3343/0.7497 & 0.3347/0.7492 & 0.2920/0.7843 & 0.3005/0.7775 & 0.3001/0.7779& 0.3398/0.7450& 0.3336/0.7500& 0.2664/0.8027&0.3401/0.7449&0.3314/0.7520\\   
 DWT-DCT \cite{DWT-DCT} &0.4703/0.6315 & 0.4696/0.6322 & 0.4026/0.6883 & 0.4480/0.6513 & 0.4826/0.6209& 0.4913/0.6133& 0.4932/0.6114& 0.1859/0.8548&0.4768/0.6264&0.4850/0.6187\\
DWT-DCT-SVD \cite{lei2016audio}& 0.3340/0.7474 & 0.3377/0.7445 & 0.2928/0.7812 &0.3077/0.7693 & 0.3422/0.7406& 0.3599/0.7257& 0.3750/0.7132& 0.3200/0.7565&0.3662/0.7205&0.3818/0.7073\\
				
   \hline
		
			\end{tabular}}
	\end{table*}

\section{Experimental results and analysis}
\label{sec:results}
\subsection{Experimental setup }
\subsubsection{Database}
In order to verify the effectiveness of the proposed zero-watermarking scheme, various styles of audio clips are randomly selected from the DSD100 database \cite{data}, including rock, classical, jazz, country and pop music. There are a total of forty pieces of music and each audio clip with a duration of 64 seconds sampled at 44.1kHz and quantized with 16 bits. A binary image of size $64 \times 64$ is adopted to generate a zero-watermark, as shown in Fig. \ref{fig:water} (a).

\subsubsection{Performance metrics}
In this work, bit error rate (BER) and normalized cross-correlation coefficient (NC) are used to evaluate the reliability of the proposed scheme and measure its anti-attack ability, respectively \cite{DWT,stft,DWT-DCT,lei2016audio}. The following common attacks and synchronization attacks are often used when evaluating robustness.

{\bf Common attacks:}
\begin{itemize}
	\item AWGN (Additive white Gaussian noise): The SNR of AWGN is 10dB and 20dB, respectively.
	\item LPF (Low pass filter): The watermarked signals are filtered through a low-pass filter with a cut-off frequency of 11,025Hz.
	\item Re-sampling: Changing the sampling frequency to 22.05kHz, and then re-sampling to 44.1kHz.
	\item Re-quantization: The number of bits is reduced from 16 bits to 8 bits, and then increased from 8 bits to 16 bits.
	\item MP3: The watermarked signals are compressed in MP3 format (128kbps).
	\item Amplitude: The amplitudes of the watermarked signals are amplified by 1.5 times and 2 times, respectively. 	
\end{itemize}

{\bf Synchronization attacks:}
\begin{itemize}
	\item TSM (Time scale modification): Modify the time scale of the watermark signal.
	\item Cropping: The front or back of the watermark signal is cropped by a few frames.
\end{itemize}

\subsection{Performance comparison under common attacks}

Table \ref{tab:attack_c} shows the performance comparison between the proposed scheme and the baselines under common attacks. We observe that both the proposed scheme and the first four schemes can effectively extract the watermark bits under the amplitude attacks, while the scheme presented in \cite{lei2016audio} is unable to resist the amplitude attacks. In addition, the robustness of the proposed scheme under the first six attacks is as superior as the scheme \cite{lei2016audio}, yet it is significantly better than the schemes based on the other three transform domains. The reason behind this result may be attributed to the robust graph Fourier coefficients and use of K-means to cluster the feature sequence to obtain the feature binary sequence. Overall, the proposed scheme has an outstanding performance in resisting common attacks and outperforms the baselines. In addition, the watermark images located in the first row of Fig. \ref{fig:water} are extracted after the host signal is subjected to common attacks. The extracted images appear almost the same as the original, which depicts that the robustness of the proposed method.

\subsection{Performance comparison under synchronization attacks}

Table \ref{tab:attack_cc} shows the performance comparison between the proposed scheme and the baselines under synchronization attacks. It can be observed that as the TSM changes larger, the robustness of the proposed  scheme decreases. This trend can also be found under cropping attacks, and the robustness of the proposed scheme decreases as the number of cropped frames increases. The reason behind this trend can be explained by the watermark extraction process. As the watermark is sequentially embedded in the host audio signal, and the host signal is subject to TSM and cropping attacks, the characteristics of the corresponding audio segment may be inappropriate, which will affect the correct extraction of the watermark. In the second row of Fig. \ref{fig:water}, the watermarked images extracted under the synchronization attacks are all slightly garbled, which is more evident in case of more serious attacks. However, it is noted that the meaningful information of the watermark images can still be obtained. Additionally, compared with the baselines, the proposed scheme shows a better robustness under TSM and cropping attacks. This may be related to the robust graph Fourier coefficients and use of K-means to cluster the feature sequence to obtain the feature binary sequence.


\section{Conclusion}
In this work, we propose a novel zero-watermarking technique based on the GFT. We note that the combined shift operator is used to construct the graph signal, and then the stable graph Fourier coefficients are selected for encoding to obtain zero-watermark embedding. Our experimental results on DSD100 database show that the proposed scheme is more robust than the traditional transform domains such as DCT, DWT, FT against common and synchronization attacks.


%

\clearpage
\balance
{
\bibliographystyle{IEEEtran}
\bibliography{BibEntries}

\begin{thebibliography}{10}
\providecommand{\url}[1]{#1}
\csname url@samestyle\endcsname
\providecommand{\newblock}{\relax}
\providecommand{\bibinfo}[2]{#2}
\providecommand{\BIBentrySTDinterwordspacing}{\spaceskip=0pt\relax}
\providecommand{\BIBentryALTinterwordstretchfactor}{4}
\providecommand{\BIBentryALTinterwordspacing}{\spaceskip=\fontdimen2\font plus
\BIBentryALTinterwordstretchfactor\fontdimen3\font minus
  \fontdimen4\font\relax}
\providecommand{\BIBforeignlanguage}[2]{{%
\expandafter\ifx\csname l@#1\endcsname\relax
\typeout{** WARNING: IEEEtran.bst: No hyphenation pattern has been}%
\typeout{** loaded for the language `#1'. Using the pattern for}%
\typeout{** the default language instead.}%
\else
\language=\csname l@#1\endcsname
\fi
#2}}
\providecommand{\BIBdecl}{\relax}
\BIBdecl

\bibitem{Non-Linear}
T.~Zong, Y.~Xiang, I.~Natgunanathan, L.~Gao, G.~Hua, and W.~Zhou,
  ``Non-linear-echo based anti-collusion mechanism for audio signals,''
  \emph{IEEE/ACM Transactions on Audio, Speech, and Language Processing},
  vol.~29, pp. 969--984, 2021.

\bibitem{blind}
P.~Dhar and T.~Shimamura, ``Blind audio watermarking in transform domain based
  on singular value decomposition and exponential-log operations,''
  \emph{Radioengineering}, vol.~26, pp. 552--561, 06 2017.

\bibitem{HE}
Y.~Xiang, I.~Natgunanathan, Y.~Rong, and S.~Guo, ``Spread spectrum-based high
  embedding capacity watermarking method for audio signals,'' \emph{IEEE/ACM
  Transactions on Audio, Speech, and Language Processing}, vol.~23, no.~12, pp.
  2228--2237, 2015.

\bibitem{PN}
Y.~Xiang, I.~Natgunanathan, D.~Peng, G.~Hua, and B.~Liu, ``Spread spectrum
  audio watermarking using multiple orthogonal pn sequences and variable
  embedding strengths and polarities,'' \emph{IEEE/ACM Transactions on Audio,
  Speech, and Language Processing}, vol.~26, no.~3, pp. 529--539, 2018.

\bibitem{8577943}
K.~Yang, W.~Wang, Z.~Yuan, and W.~Zhao, ``Strong robust zero watermarking
  algorithm based on nsct transform and image normalization,'' in \emph{2018
  IEEE 3rd Advanced Information Technology, Electronic and Automation Control
  Conference (IAEAC)}, 2018, pp. 236--240.

\bibitem{V-map}
Y.~Peng and M.~Yue, ``A zero-watermarking scheme for vector map based on
  feature vertex distance ratio,'' \emph{Journal of Electrical and Computer
  Engineering}, vol. 2015, 01 2015.

\bibitem{CONTENT}
M.~Cao, C.~Li, and L.~Tian, ``Content-based audio zero-watermarking algorithm
  against tsm,'' in \emph{2016 5th International Conference on Informatics,
  Electronics and Vision (ICIEV)}, 2016, pp. 297--301.

\bibitem{beats}
Y.~Sun, L.~Tian, and C.~Li, ``Robust zero-watermarking algorithm based on audio
  beats,'' in \emph{2021 6th International Conference on Intelligent Computing
  and Signal Processing (ICSP)}, 2021, pp. 308--312.

\bibitem{DCT--HT}
A.~Kanhe and A.~Gnanasekaran, ``A blind audio watermarking scheme employing
  dct--ht--sd technique,'' \emph{Circuits, Systems, and Signal Processing},
  vol.~38, no.~8, pp. 3697--3714, 2019.

\bibitem{imperceptible}
H.~Karajeh and M.~Maqableh, ``An imperceptible, robust, and high payload
  capacity audio watermarking scheme based on the dct transformation and schur
  decomposition,'' \emph{Analog Integrated Circuits and Signal Processing},
  vol.~99, no.~3, pp. 571--583, 2019.

\bibitem{MP3}
K.~Wang, C.~Li, and L.~Tian, ``Audio zero watermarking for mp3 based on low
  frequency energy,'' in \emph{2017 6th International Conference on
  Informatics, Electronics and Vision \& 2017 7th International Symposium in
  Computational Medical and Health Technology (ICIEV-ISCMHT)}.\hskip 1em plus
  0.5em minus 0.4em\relax IEEE, 2017, pp. 1--5.

\bibitem{DWT}
Y.~Yang, M.~Lei, M.~Cheng, B.~Liu, G.~Lin, and D.~Xiao, ``An audio
  zero-watermark scheme based on energy comparing,'' \emph{China
  Communications}, vol.~11, no.~7, pp. 110--116, 2014.

\bibitem{DWT-dsss}
S.~Choudhary, K.~Nath, and J.~Panda, ``Double layered audio zero-watermarking
  using dwt \& dsss,'' in \emph{2017 International Conference on Communication
  and Signal Processing (ICCSP)}.\hskip 1em plus 0.5em minus 0.4em\relax IEEE,
  2017, pp. 0419--0423.

\bibitem{dft}
R.~Subhashini and K.~B. Bagan, ``Robust audio watermarking for monitoring and
  information embedding,'' in \emph{2017 Fourth International Conference on
  Signal Processing, Communication and Networking (ICSCN)}.\hskip 1em plus
  0.5em minus 0.4em\relax IEEE, 2017, pp. 1--4.

\bibitem{stft}
A.~E.~A. Jayarani, M.~R. Bhatt, and D.~Geetha, ``Zero watermarking on audio
  based on stft,'' in \emph{2018 International Conference on Computing,
  Electronics \& Communications Engineering (iCCECE)}.\hskip 1em plus 0.5em
  minus 0.4em\relax IEEE, 2018, pp. 253--256.

\bibitem{lpcc}
S.-M. Tsai, ``An efficient and robust zero-watermarking scheme for digital
  audio,'' in \emph{2013 IEEE International Conference on Circuits and Systems
  (ICCAS)}.\hskip 1em plus 0.5em minus 0.4em\relax IEEE, 2013, pp. 51--54.

\bibitem{dhar2017blind}
P.~K. Dhar and T.~Shimamura, ``Blind audio watermarking in transform domain
  based on singular value decomposition and exponential-log operations,''
  \emph{Radioengineering}, vol.~26, no.~2, pp. 552--561, 2017.

\bibitem{DWT-DCT}
J.~Panda, S.~Choudhary, K.~Nath, and S.~Kumar, ``Audio zero watermarking scheme
  based on sub band mean energy comparison using dwt-dct,'' in \emph{2016
  International Conference on Signal Processing and Communication
  (ICSC)}.\hskip 1em plus 0.5em minus 0.4em\relax IEEE, 2016, pp. 352--357.

\bibitem{lei2016audio}
M.~Lei, Y.~Yang, X.~Liu, M.~Cheng, and R.~Wang, ``Audio zero-watermark scheme
  based on discrete cosine transform-discrete wavelet transform-singular value
  decomposition,'' \emph{China Communications}, vol.~13, no.~7, pp. 117--121,
  2016.

\bibitem{gft}
\BIBentryALTinterwordspacing
A.~Sandryhaila and J.~Moura, ``Discrete signal processing on graphs: Frequency
  analysis,'' \emph{IEEE Transactions on Signal Processing}, vol.~62, no.~12,
  pp. 3042--3054, 2014, cited By 396. [Online]. Available:
  \url{https://www.scopus.com/inward/record.uri?eid=2-s2.0-84901346744&doi=10.1109%2fTSP.2014.2321121&partnerID=40&md5=b2ef052da487870ebe44011adbabbc94}
\BIBentrySTDinterwordspacing

\bibitem{limit}
A.~Ortega, P.~Frossard, J.~Kovačević, J.~M.~F. Moura, and P.~Vandergheynst,
  ``Graph signal processing: Overview, challenges, and applications,''
  \emph{Proceedings of the IEEE}, vol. 106, no.~5, pp. 808--828, 2018.

\bibitem{graph}
D.~I. Shuman, S.~K. Narang, P.~Frossard, A.~Ortega, and P.~Vandergheynst, ``The
  emerging field of signal processing on graphs: Extending high-dimensional
  data analysis to networks and other irregular domains,'' \emph{IEEE Signal
  Processing Magazine}, vol.~30, no.~3, pp. 83--98, 2013.

\bibitem{gss}
\BIBentryALTinterwordspacing
X.~Yan, Z.~Yang, T.~Wang, and H.~Guo, ``An iterative graph spectral subtraction
  method for speech enhancement,'' \emph{Speech Communication}, vol. 123, pp.
  35--42, 2020. [Online]. Available:
  \url{https://www.sciencedirect.com/science/article/pii/S0167639320302405}
\BIBentrySTDinterwordspacing

\bibitem{3D}
\BIBentryALTinterwordspacing
E.~E. Abdallah, A.~B. Hamza, and P.~Bhattacharya, ``Spectral graph-theoretic
  approach to 3d mesh watermarking,'' in \emph{Proceedings of Graphics
  Interface 2007}, ser. GI '07.\hskip 1em plus 0.5em minus 0.4em\relax New
  York, NY, USA: Association for Computing Machinery, 2007, p. 327–334.
  [Online]. Available: \url{https://doi.org/10.1145/1268517.1268570}
\BIBentrySTDinterwordspacing

\bibitem{GS}
H.~Al-Khafaji and C.~Abhayaratne, ``Graph spectral domain blind watermarking,''
  in \emph{ICASSP 2019 - 2019 IEEE International Conference on Acoustics,
  Speech and Signal Processing (ICASSP)}, 2019, pp. 2492--2496.

\bibitem{2020A}
F.~Ferreira and J.~B. Lima, ``A robust 3d point cloud watermarking method based
  on the graph fourier transform,'' \emph{Multimedia Tools and Applications},
  vol.~79, no.~1, pp. 1--30, 2020.

\bibitem{sensor}
H.~Al-khafaji and C.~Abhayaratne, ``Graph spectral domain watermarking for
  unstructured data from sensor networks,'' in \emph{2017 22nd International
  Conference on Digital Signal Processing (DSP)}, 2017, pp. 1--5.

\bibitem{L2016}
E.~Pavez and A.~Ortega, ``Generalized laplacian precision matrix estimation for
  graph signal processing,'' in \emph{2016 IEEE International Conference on
  Acoustics, Speech and Signal Processing (ICASSP)}, 2016, pp. 6350--6354.

\bibitem{A2018}
A.~Hiruma, K.~Yatabe, and Y.~Oikawa, ``Separating stereo audio mixture having
  no phase difference by convex clustering and disjointness map,'' in
  \emph{2018 16th International Workshop on Acoustic Signal Enhancement
  (IWAENC)}, 2018, pp. 266--270.

\bibitem{data}
A.~Liutkus, F.-R. St{\"o}ter, Z.~Rafii, D.~Kitamura, B.~Rivet, N.~Ito, N.~Ono,
  and J.~Fontecave, ``The 2016 signal separation evaluation campaign,'' in
  \emph{Latent Variable Analysis and Signal Separation - 12th International
  Conference, {LVA/ICA} 2015, Liberec, Czech Republic, August 25-28, 2015,
  Proceedings}, P.~Tichavsk{\'y}, M.~Babaie-Zadeh, O.~J. Michel, and
  N.~Thirion-Moreau, Eds.\hskip 1em plus 0.5em minus 0.4em\relax Cham: Springer
  International Publishing, 2017, pp. 323--332.

\end{thebibliography}
}

\end{document}